\documentclass[preprint,showpacs,preprintnumbers,amsmath,amssymb]{revtex4}

\usepackage{graphicx}
\usepackage{dcolumn}
\usepackage{bm}

\begin{document}
\def\brho{{\hbox{\boldmath $\rho$}}}
\def\bsb{{\hbox{\boldmath $\beta$}}}
\def\bsk{{\hbox{\boldmath $k$}}}
\def\bsp{{\hbox{\boldmath $p$}}}

\title{Imaging of granular sources in high energy heavy ion collisions}

\author{Zhi-Tao Yang$^1$}
\author{Wei-Ning Zhang$^{1,2}$\footnote{wnzhang@dlut.edu.cn}}
\author{Lei Huo$^{1}$}
\author{Jing-Bo Zhang$^{1}$}

\affiliation{
$^1$Department of Physics, Harbin Institute of
Technology, Harbin, Heilongjiang 150006, China\\
$^2$School of Physics and Optoelectronic Technology, Dalian
University of Technology, Dalian, Liaoning 116024, China
}

\date{\today}

\begin{abstract}
We investigate the source imaging for a granular pion-emitting
source model in high energy heavy ion collisions.  The two-pion
source functions of the granular sources exhibit a two-tiered
structure.  Using a parametrized formula of granular two-pion source
function, we examine the two-tiered structure of the source
functions for the imaging data of Au+Au collisions at Alternating
Gradient Synchrotron (AGS) and Relativistic Heavy Ion Collider
(RHIC).  We find that the imaging technique introduced by Brown and
Danielewicz is suitable for probing the granular structure of the
sources.  Our data-fitting results indicate that there is not
visible granularity for the sources at AGS energies. However, the
data for the RHIC collisions with the selections of $40 < {\rm
centrality} < 90\%$ and $0.20<k_{\rm T}<0.36$ GeV/c are better
described by the model with granular emission than that of one
Gaussian.  The model with granular source has more parameters than
the simple Gaussian, hence can describe more complicated shapes.
\end{abstract}

\pacs{25.75.-q, 25.75.Nq, 25.75.Gz}

\maketitle

\section{Introduction}

Two-pion Hanbury-Brown-Twiss (HBT) interferometry is a powerful tool
of detection of the space-time structure of particle-emitting
sources produced in high energy heavy ion collisions
\cite{Won94,Wie99,Wei00,Lis05}.  In conventional two-pion HBT
analysis one needs fitting the two-pion HBT correlation functions
with parametrized formulas to obtain quantitatively the source
space-time results.  So the explanations of the HBT results are
model depended.  Imaging technique introduced by Brown and
Danielewicz \cite{Bro97,Bro98,Bro01} is a model-independent way.  It
can be used to extract the source geometry pictures (source
function) directly from the two-pion correlation functions.  This
technique has been developed and used in high energy heavy ion
collisions
\cite{Pan99,Bro00,E89501,Ver02,E89503,Dan04,Chu05,Bro05,Dan07,PHE07,Lac07,Afa07,Lac08}.

Recently, there has been much progress in understanding of the
process of nucleus-nucleus collisions at RHIC.  However, there are
still many unsolved problems. One of them is the so-called HBT
puzzle, $R_{\rm out} / R_{\rm side} \approx 1$
\cite{STA01a,PHE02a,PHE04a,STA05a}.  Here $R_{\rm out}$ and $R_{\rm
side}$ are the transverse HBT radii parallel and perpendicular to
the pion pair momentum \cite{Pra86,Ber88}.  In Ref. \cite{Zha04} a
granular source model of quark-gluon plasma (QGP) droplets was put
forth to explain the HBT puzzle.  The suggestion was based on the
observation that in the hydrodynamic calculations for the granular
source the average particle emission time scales with the initial
radius of the droplet, whereas the spacial size of the source is the
scale of the distribution of the droplets.  For a granular source
with many of the small droplets distributed in a relatively large
region, the HBT radius $R_{\rm out}$ can be close to $R_{\rm side}$
\cite{Zha04}. In Ref. \cite{Zha06} the authors further investigated
the elliptic flow and HBT radii as a function of pion transverse
momentum for an improved granular source model of QGP droplets. They
argued that although a granular structure was suggested earlier as
the signature of a first-order phase transition
\cite{Wit84,Pra92,Cse92,Ala99,Ran04,Zha95,Zha00,Zha04,Won04,Zha05},
the occurrence of granular structure may not be limited to
first-order phase transition \cite{Zha06,Zha07}.  The large
fluctuations of initial matter distribution
\cite{Gyu97,Dre02,Ham04,And08} in high energy heavy ion collisions
may facilitate the occurrence of instability of the system during
its subsequent expansion and fragmentation to many granular droplets
together with surface tension effect \cite{Zha06,Zha07}.

Recent researches on event-by-event two-pion Bose-Einstein
correlations in smoothed particle hydrodynamics indicate that the
particle-emitting sources produced at RHIC energy are inhomogeneous
and there is a granular structure of many ``lumps" (droplets)
\cite{Ren08}.  For a simple granular source model we will show that
the two-pion source function has a two-tiered structure.  In small
relative coordinate $r$ region, the source function exhibits an
enhancement because of the higher density in the droplets.  Previous
RHIC experimental imaging researches are mainly focused on the
long-range tail of the source functions at large $r$
\cite{PHE07,Lac07,Afa07,Lac08}, and the deviations of the source
function from Gaussian distribution in the large $r$ region are
believed mainly the contribution of long-lived resonances
\cite{PHE07,Bro07}.  In this paper we will focus our attention on
the source functions in small $r$ region.  We will investigate the
imaging of granular sources.  We will examine the two-tiered
structure of the source functions for the AGS and RHIC imaging data
of Au+Au \cite{E89501,PHE07}.  Our results indicate that the imaging
technique is suitable for probing the granular structure of the
particle-emitting sources.  There is not visible granularity for the
sources in Au+Au collisions at AGS energies.  However, the data for
the RHIC collisions with the selections of $40 < {\rm centrality} <
90\%$ and $0.20<k_{\rm T}<0.36$ GeV/c are better described by the
model with granular emission than from that of one Gaussian.

\section{Imaging technique}

For the convenience of discussion later in the paper, we start out
with a brief review of the imaging technique of Brown and
Danielewicz \cite{Bro97,Bro98,Bro01}.

Based on the Koonin-Pratt formulism \cite{Koo77,Pra90}, the
two-pion HBT correlation function may be expressed in the
center-of-mass frame of the particle pair as
\cite{Bro01,Bro00,E89501,Bro05}:
\begin{eqnarray}
\label{cf1} C({\bf q})-1=\int\! d{\bf r}K({\bf q},{\bf r})S({\bf
r}),
\end{eqnarray}
where ${\bf q}={\bf p}_1 - {\bf p}_2$ is the relative momentum of
the pion pair, $\bf r$ is the relative separation of emission points
of the two particles, $K({\bf q},{\bf r})=|\Phi_{\bf q}({\bf
r})|^2-1$, where $\Phi_{\bf q}({\bf r})$ is the relative wave
function of the pair. Neglecting the final-state interaction of the
pion pair, one has
\begin{equation}
\Phi_{\bf q}({\bf r})=\frac{1}{\sqrt{2}}(e^{i{\bf q}\cdot{\bf r}/2}
+ e^{-i{\bf q}\cdot{\bf r}/2})\,.
\end{equation}
In Eq. (\ref{cf1}), $S({\bf r})$ is the so-called two-particle
source function.  It may be written with Wigner function as
\cite{Bro01,Bro00},
\begin{eqnarray}
\label{Wigner} S({\bf r}) &=& \int\!dt \int\!d^3R\,dT\,D({\bf
R}\!+\!{\bf r}/2, T\!+\!t/2, {\bf p}_1 )\nonumber\\
&&\times D({\bf R}\!-\!{\bf r}/2, T\!-\!t/2, {\bf p}_2 )\,.
\end{eqnarray}
Here the Wigner functions are normalized particle emission rates,
\begin{equation}
D({\bf r}, t, {\bf p})=\frac{E\,d^7N}{d^3r\, dt\, d^3p} \bigg / \int
\frac{E\,d^3N}{d^3p} d^3p\,.
\end{equation}

For a spherically symmetric source function, $S({\bf r})=S(r)$,
performing the angle integrations on the right hand of Eq.
(\ref{cf1}), one gets the angle-averaged version of Eq. (\ref{cf1})
as,
\begin{eqnarray}
\label{cf2} {\cal R}(q_{\rm inv}) \equiv C(q_{\rm inv})-1=4\pi\int\!
dr\, r^2 K(q_{\rm inv},r) S(r).
\end{eqnarray}
Here $q_{\rm inv}=\sqrt{{\bf q}^2 - q_0^2}$,
\begin{eqnarray}
K(q_{\rm inv},r)=\sin(q_{\rm inv} r)/(q_{\rm inv} r)\,.
\end{eqnarray}
Equation (\ref{cf2}) now is suitable in any frame and the problem of
imaging becomes inverting $K(q_{\rm inv},r)$ with measured
correlation function ${\cal R}(q_{\rm inv})$.  After expanding the
source function $S(r)$ in $b-{\rm spline}$ basis \cite{Bro01},
\begin{eqnarray}
S(r)=\sum_j S_j B_j(r),
\end{eqnarray}
the inversion problem reduces to the problem of solving the matrix
equation
\begin{eqnarray}
\label{Rqi}
{\cal R}(q_i)=\sum_j K_{ij} S_j,
\end{eqnarray}
where ${\cal R}(q_i)$ denotes the value of the correlation function
at the $i$th bin of $q_{\rm inv}$, and
\begin{eqnarray}
K_{ij}=\frac{4\pi}{\Delta q}\int_{q_i-\Delta q/2}^{q_i+\Delta
q/2}dq_{\rm inv} \int_0^\infty dr\, r^2 K(q_{\rm inv}, r) B_j(r),
\end{eqnarray}
where $\Delta q$ is the bin size of $q_{\rm inv}$.

In present paper, the minimization package MINUIT \cite{MINUIT} was
used to minimize the $\chi^2$ between the measured and calculated
correlation functions.

\section{Source function imaging for static granular sources}

In this section we examine the source function imaging for static
granular sources.  Although static source is not a realistic case,
its source function may still be used as a parametrized formula in
the analysis for evolving sources.

For the granular source model, particles are emitted from dispersed
droplets \cite{Pra92,Zha95,Zha00,Zha04,Won04,Zha05,Zha06}.  Assuming
that the granular source has the same $N$ droplets and the
distribution of the particle emission points in a droplet has
Gaussian form, the normalized source density distribution is given
by \cite{Won04}
\begin{eqnarray}
\label{bkl2} D({\bf x})=\frac{1}{N(\sqrt{2\pi}\, a)^3}\sum_{i=1}^N
\exp\big[-\frac{({\bf x}-{\bf X}_i) ^2}{2\,a^2}\big]\,,
\end{eqnarray}
where $a$ is the ``radius" of the droplets and ${\bf X}_i$ is the
spatial coordinate of the $i$th droplet center.  Inserting this
distribution into Eq. (\ref{Wigner}), we obtain
\begin{eqnarray}
\label{gs1}
S({\bf r})&=&\frac{1}{N^2(\sqrt{4\pi}a)^3}\sum_{i,j=1}^N
\exp\left[-\frac{({\bf r}-{\bf X}_{ij})^2}{4\,a^2}
\right] \notag\\
&=&\frac{1}{N^2(\sqrt{4\pi}
a)^3}\exp\left(-\frac{r^2}{4\,a^2}\right)\notag\\
&\times& \!\!\sum_{i,j=1}^N\exp\left(-\frac{|{\bf
X}_{ij}|^2}{4\,a^2}\right) \exp\left(\frac{r|{\bf
X}_{ij}|\cos\alpha}{2\,a^2}\right),~~~~
\end{eqnarray}
where ${\bf X}_{ij}={\bf X}_i-{\bf X}_j$, $\alpha$ is the angle
between ${\bf r}$ and ${\bf X}_{ij}$.  The source function presents
fluctuation due to the factor $\exp\left(\frac{r|{\bf
X}_{ij}|\cos\alpha}{2\,a^2}\right)$.

Due to limited number of produced particles per event, conventional
two-pion HBT interferometry are analyses based on averages over
events.  The correlation functions are obtained from the correlated
pion pairs (the two identical pions in each of the pairs are from
the same event) of all sample events. Accordingly, the source
function for the mixed-events of the granular source, $S^{\rm
Gran}(r)$, should be the average of Eq. (\ref{gs1}) over all the
events. Assuming that the droplet centers in the granular source
obey the Gaussian distribution, $P({\bf X}_i) \sim \exp(-{\bf
X}_i^2/2R_{\rm gr}^2)$, we get
\begin{eqnarray}
\label{SGranu} &&S^{\rm Granu}(r)
=\frac{1}{N}\frac{1}{(\sqrt{4\pi}a)^3}\exp\Big (-\frac{
r^2}{4\,a^2}\Big ) + \Big (1-\frac{1}{N}\Big)\nonumber \\
&&~~~~\times \frac{1}{(\sqrt{4\pi}\sqrt{a^2+R_{\rm gr}^2})^3}
\exp\bigg[-\frac{r^2}{4(a^2+R_{\rm gr}^2)}\bigg].~~~~~~~~
\end{eqnarray}
Compared with the two-particle source function of the Gaussian
source model usually used \cite{Bro01,E89501,Bro05},
\begin{equation}
\label{SGauss}
S^{\rm Gauss}(r) = \frac{1}{(\sqrt{4\pi} R_{\rm ga})^3}
\exp\big(-\frac{r^2}{4R^2_{\rm ga}}\big),
\end{equation}
the source function of the granular source consists of two
exponential terms, which correspond to that the two particles from
the same droplet and from different droplets respectively.  This
``two-tiered structure" of the source function is consistent with
the two-tiered structure of the correlation function for the
granular source \cite{Pra92,Zha95}. It is most apparent when $N=2$,
and disappears when $N \to \infty$.  For $R_{\rm gr}^2 \gg a^2$, the
coefficient ratio of the two terms in Eq. (\ref{SGranu}) is $(R_{\rm
gr}/a)^3/(N-1)$, which is a characteristic quantity for the
two-tiered structure.  Considering also $N \ge 2$ for the granular
source model, we introduce the quantity,
\begin{equation}
\xi=\frac{(R_{\rm gr}/a)^3}{N-2},
\end{equation}
to characterize the granularity of the sources.  For granular
sources the values of $\xi$ are between $(0,\infty)$, and a source
will have not granularity when $\xi \le 0$.  In Fig. 1 we show the
curves of $S^{\rm Granu}(r)$ as a function of the droplet number $N$
for the granular sources with $R_{\rm gr}=4.5$ fm and $a=1.5$ fm.
One can see that the the two-tiered structure is obvious for finite
droplet numbers of the granular sources.

\begin{figure}
\includegraphics[angle=0,scale=0.50]{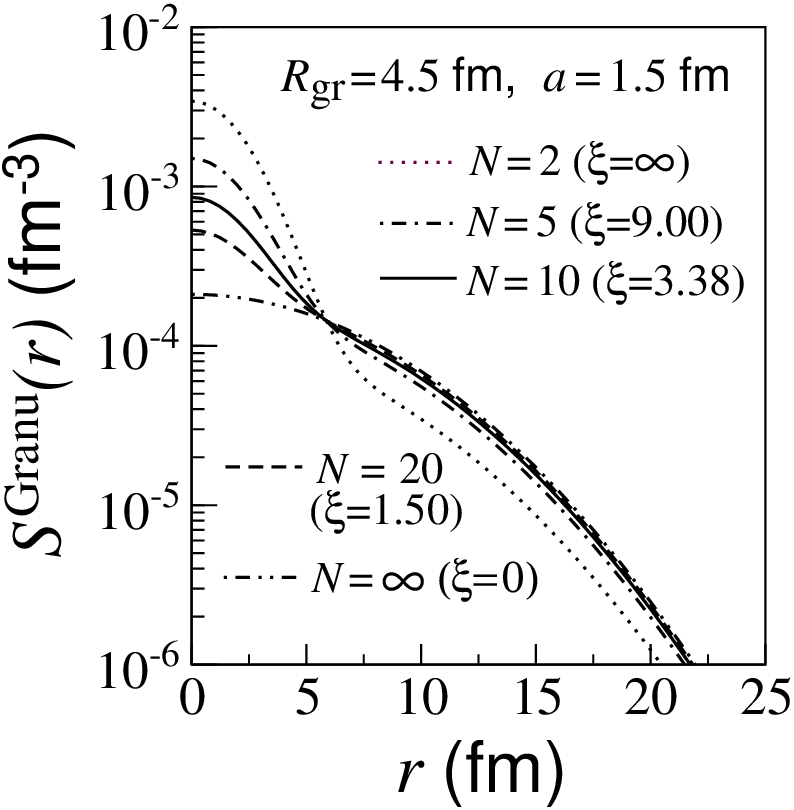}
\caption{\label{fig:fig1} The two-particle source functions of
granular sources.}
\end{figure}

We next examine the imaging of the static granular source.  In Fig.
2(a) we show the two-pion correlation function (CF) obtained from
$2\times10^5$ simulated two-pion events for the granular source with
$R_{\rm gr}=4.5$ fm and $a=1.5$ fm.  For comparison, the correlation
function for a Gaussian source with $R_{\rm ga}=4.5$ fm is presented
in Fig. 2(b).  In the simulations, pions are emitted thermally from
the sources at freeze-out temperature $T_f=150$ MeV.  Figure 2(a$'$)
and (b$'$) show the two-pion source functions (circle symbols)
extracted from the two-pion correlation functions by the imaging
technique.  The curves in Fig. 2(a$'$) and (b$'$) are the results of
the source function fit (SFF) with the formulas $\lambda S\,^{\rm
Granu}(r)$ and $\lambda S\,^{\rm Gauss}(r)$.  Here $\lambda$ is the
parameter of source coherent factor in HBT interferometry
\cite{Won94,Wie99,Wei00,Lis05} and $S\,^{\rm Granu}(r)$ and
$S\,^{\rm Gauss}(r)$ are given by Eqs. (\ref{SGranu}) and
(\ref{SGauss}), respectively.  The corresponding fitting results are
also presented in the figure.  For the granular source the
$\chi^2/{\rm NDF}$ for the granular SFF is 0.22, which is much
smaller than that of 3.88 for the Gaussian SFF.  The curves in Fig.
2(a) and (b) are the restored correlation functions (RCF) calculated
by Eq. (\ref{cf2}).  From Fig. 2 one can see that although the
two-pion correlation functions for the granular and Gaussian sources
are almost the same in shape, the granular source function exhibits
a clear enhancement in small $r$ region, which reflects the higher
source density in a droplet.  So imaging technique is suitable for
probing the granular structure of the sources.

\begin{figure}
\includegraphics[angle=0,scale=0.38]{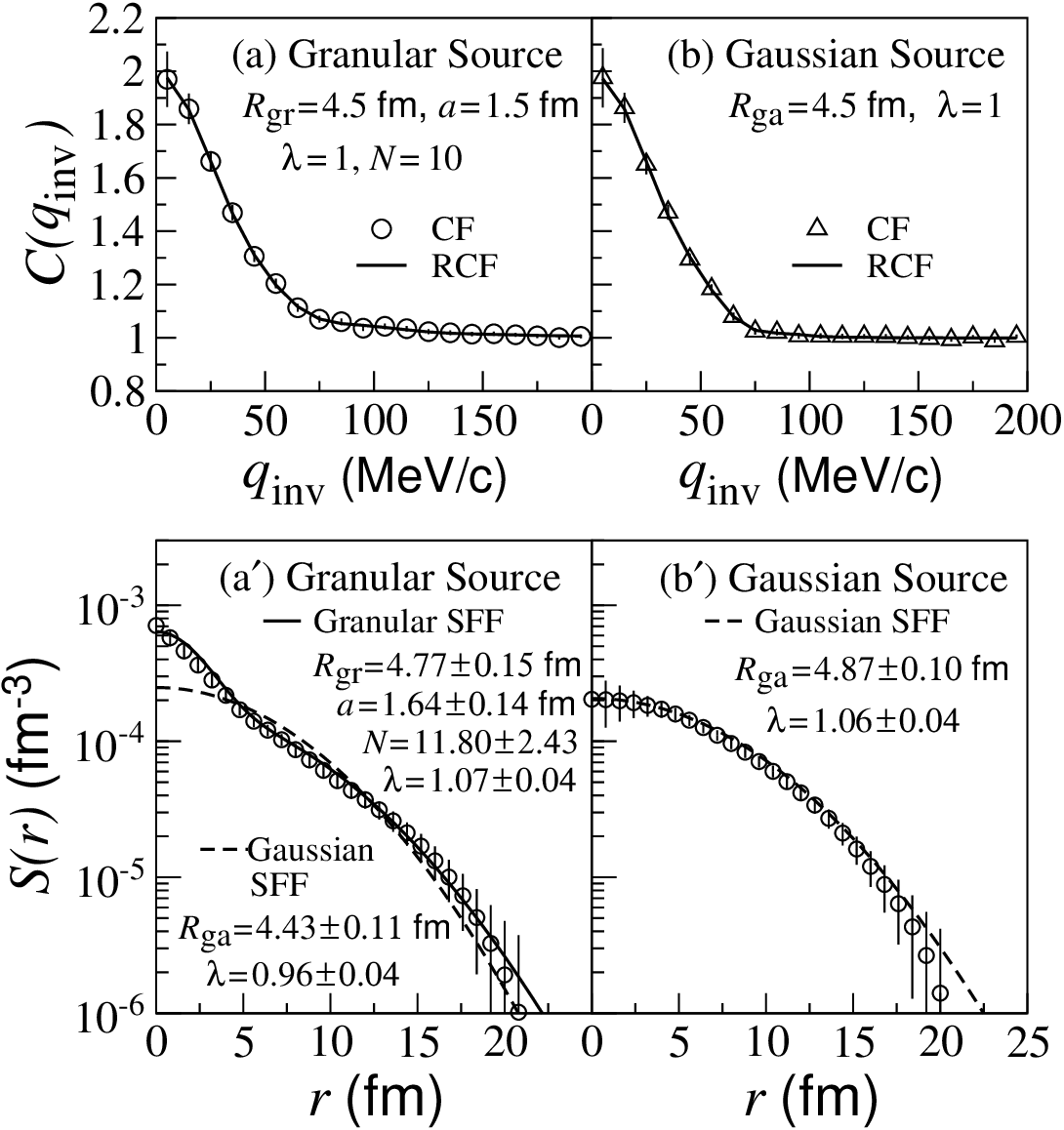}
\caption{\label{fig:fig2} (a, b) The two-pion correlation functions
for granular and Gaussian sources.  (a$'$, b$'$) The two-pion source
functions extracted by imaging technique for the granular and
Gaussian sources.}
\end{figure}

Figure 3(a), (b), (c), and (d) show further the source functions for
the granular sources with various source parameters.  It can be seen
that the two-tiered structure is more obvious for smaller droplet
number $N$ and smaller droplet radius $a$.  One can still observe
the two-tiered structure even $\xi \approx 2$.

\begin{figure}
\vspace*{5mm}
\includegraphics[angle=0,scale=0.38]{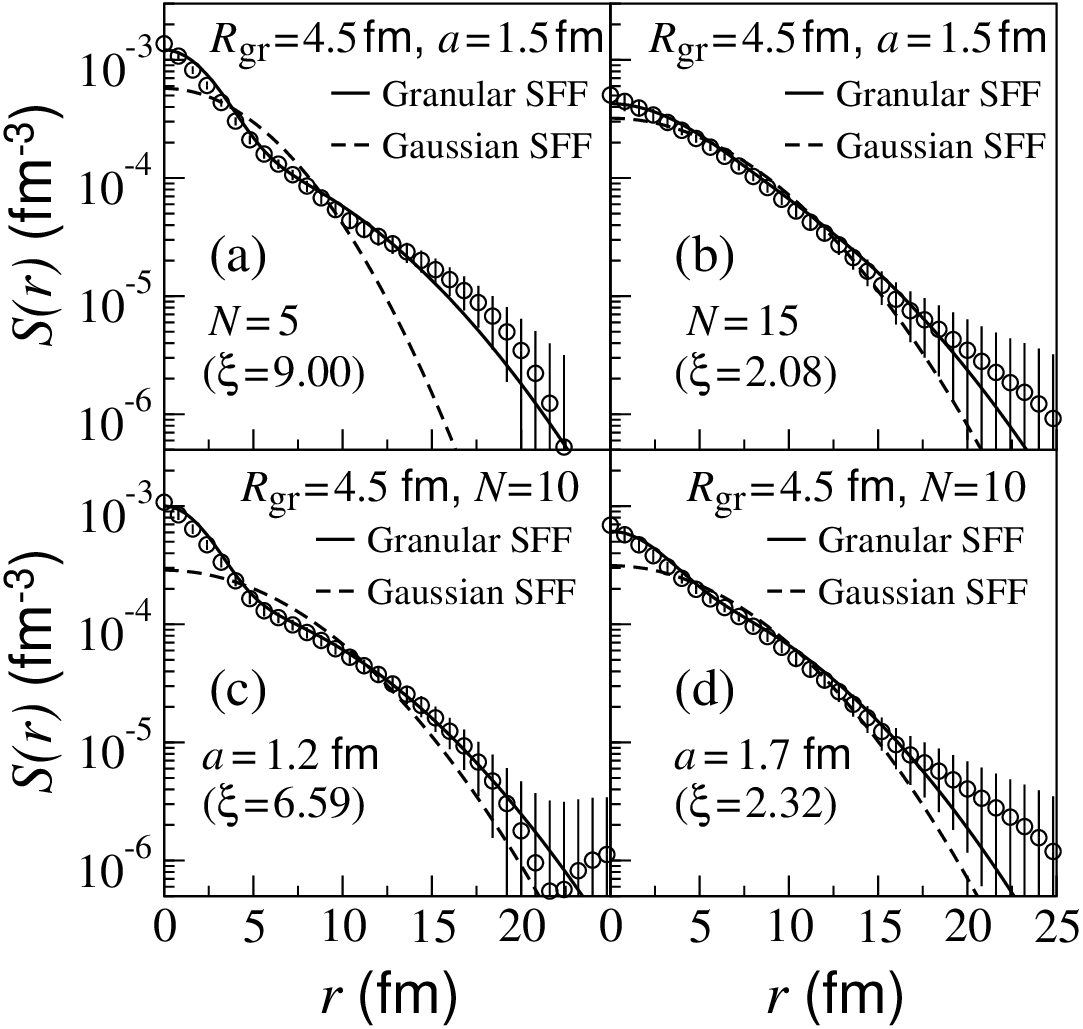}
\caption{\label{fig:fig3} The source functions for the granular
sources with various parameters.}
\end{figure}

\section{Two-tiered structure of source functions for evolving sources}

In this section we investigate the two-tiered structure of the
source functions for evolving sources.  First we consider a simple
granular source model of $N$ evolving QGP droplets.  We assume that
all of the droplets in the source have the same initial conditions
and evolve hydrodynamically in the same way \cite{Zha04,Zha05}.  An
equation of state of the entropy density suggested by QCD lattice
gauge results \cite{Bla87,Lae96,Ris96} with the transition
temperature $T_c=165$ MeV and the temperature width of the
transition $\Delta T=0.05T_c$ is used in the hydrodynamical
calculations \cite{Zha04,Zha05,Zha06}. In our calculations the
initial energy density of the droplets is taken to be
$\epsilon_0=3.75\,T_cs_c$, which is about two times of the density
of quark matter at $T_c$ \cite{Ris96,Zha06}.  The initial
distribution of the droplet centers is given by a Gaussian
distribution with standard deviation $R_0$.  For the case with an
additional collective radial expansion, the droplet centers are
assumed to have a constant radial velocity $v_d$ in the
center-of-mass frame of the granular source \cite{Zha04,Zha05}.  The
source freeze-out temperature is taken to be $T_f=150$ MeV.

Figure 4(a) and (b) show the source functions $S(r)$ (circle
symbols) obtained by the imaging technique from the two-pion
correlation functions $C(q_{\rm inv})$ for the granular sources with
the parameters $R_0=5.0$ fm, $v_d=0.5$, and $\lambda=1$.  The
droplet number and initial droplet radius for the granular source of
Fig. 4(a) are 5 and 2.5 fm, and they are 15 and 1.5 fm for the
granular source of Fig. 4(b).  One can see that the source functions
of the granular sources have enhancements in small $r$ region.  With
the granular SFF results one gets that the values of $\xi$ for the
two granular sources are $4.85\pm2.94$ and $4.66\pm2.83$,
respectively. The large $\xi$ values indicate large granularity for
the sources, which is consistent with the observations.  For
comparison, in Fig. 4(c) and (d) we show the source functions
obtained by the imaging technique from the two-pion correlation
functions for the sources with Gaussian distribution and additional
radial expanding velocities 0.3 and 0.6, respectively.  The standard
deviation for the Gaussian distribution is taken to be 5.0 fm.  From
Fig. 4(c) it can be seen that for the Gaussian distribution source
with smaller expanding velocity the granular and Gaussian SFF curves
are almost overlapped.  The granular SFF gives a very large $N$. For
a large $N$ one can see that the first term of the fitting formula
Eq. (\ref{SGranu}) approaches zero and the two-tiered structure
disappears.  One can also see from Eq. (\ref{SGranu}) that in this
case the fitting formula is almost independent of $N$ and the
parameters $R_{\rm gr}$ and $a$ can be reduced to one parameter
$\sqrt{R^2_{\rm gr}+a^2}$ as $R_{\rm ga}$ in Eq. (\ref{SGauss}).  So
the fit is insensitive to the parameters $N$ and $R_{\rm gr}$. From
Fig. 4(d) one can see that in small $r$ region the two SFF curves
are almost overlapped and there is only a small difference between
the two SFF curves in large $r$ region.  We find that the granular
SFF result of $N$ is less than 2.  In this case the source has not
granularity.  With the granular SFF results for the two expanding
Gaussian sources we get that the values of the characteristic
quantity of granularity $\xi$ are $4.813 \times 10^{-6} \pm 0.002$
and $-12.900\pm 4.185$.  They indicate that there is not granularity
for the sources.

\begin{figure}
\includegraphics[angle=0,scale=0.55]{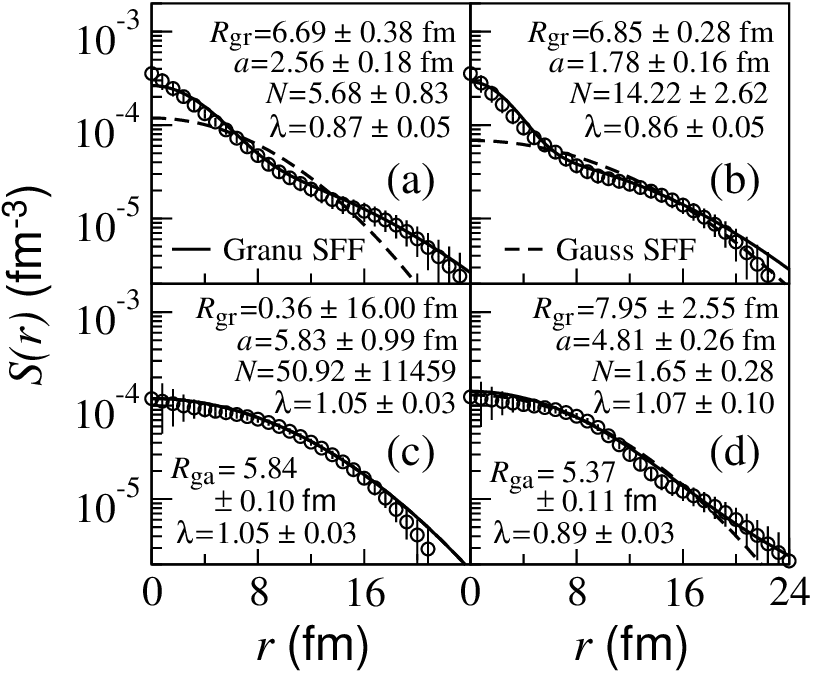}
\caption{\label{fig:fig4} The source functions for the evolving
granular sources [(a) and (b)] and the expanding Gaussian sources
[(c) and (d)].}
\end{figure}

We next examine the two-tiered structure of the two-pion source
functions for the imaging data in the Au+Au collisions at AGS
\cite{E89501} and RHIC \cite{PHE07}.  The circle symbols in Fig.
5(a), (b), (c), and (d) show the two-pion source functions for 2, 4,
6, and 8 $A$GeV Au+Au collisions, respectively \cite{E89501}. The
circle symbols in Fig. 6(a) and (b) show the source functions for
$\sqrt{s_{NN}}=200$ GeV Au+Au collisions with different cut
conditions of centrality and average transverse momentum $k_{\rm T}$
of the pion pair.  In Figs. 5 and 6, the solid and dashed curves are
our granular and Gaussian SFF curves.  The fitting results are
listed in Table I.  It can be seen that the $\chi^2\!/{\rm NDF}$ for
the granular SFF are very small.  This indicates that the data
errors are large for the granular SFF and the granular SFF may
distinguish more complicated source shapes if there is enough
statistics.  From Fig. 5 one cannot observe the two-tiered structure
of the source functions in small $r$ region. When we use the
granular parametrized formula $\lambda S^{\rm Granu}(r)$ to fit the
source functions we find that the errors of $N$ are the same order
of the values of $N$. In Table I we present the fit results for
fixed $N=50$.  One can see that the values of $\xi$ in Table I for
the collisions at AGS energies are very small. So there is not
visible granularity for the sources.

\begin{figure}
\vspace*{5mm}
\includegraphics[angle=0,scale=0.48]{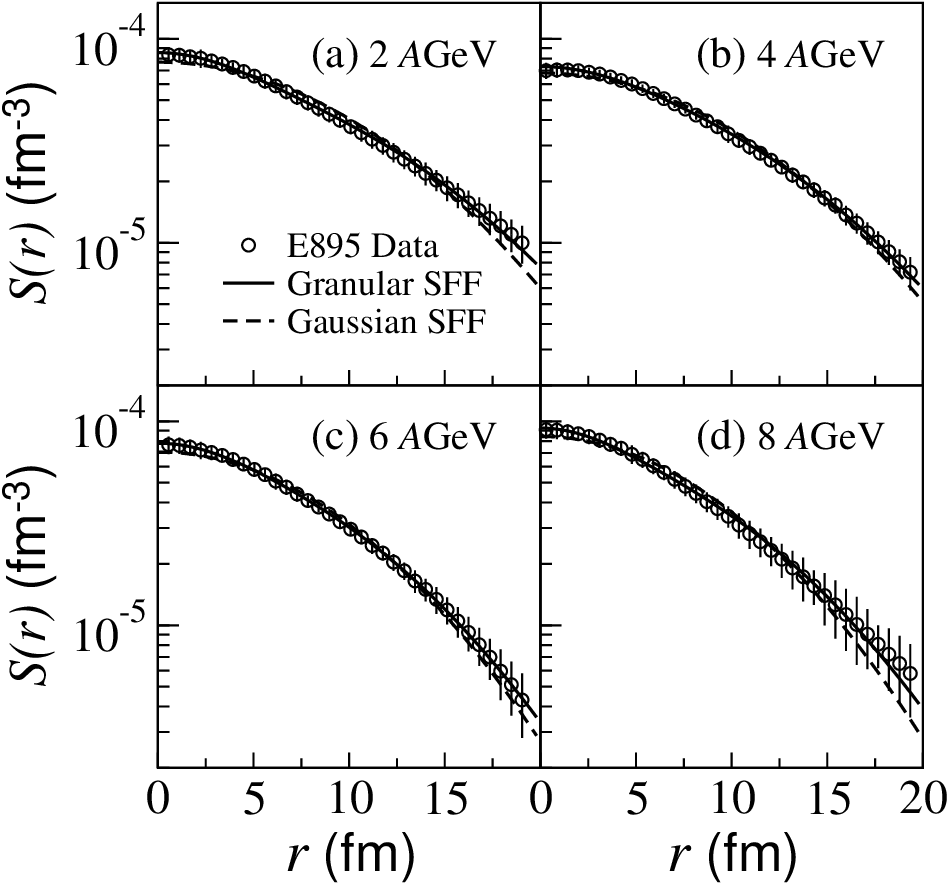}
\caption{\label{fig:fig5} The two-pion source functions ($\circ$
symbols) for 2, 4, 6, and 8 $A$GeV Au+Au collisions \cite{E89501}
and the SFF curves.}
\end{figure}

\begin{figure}
\vspace*{10mm}
\includegraphics[angle=0,scale=0.34]{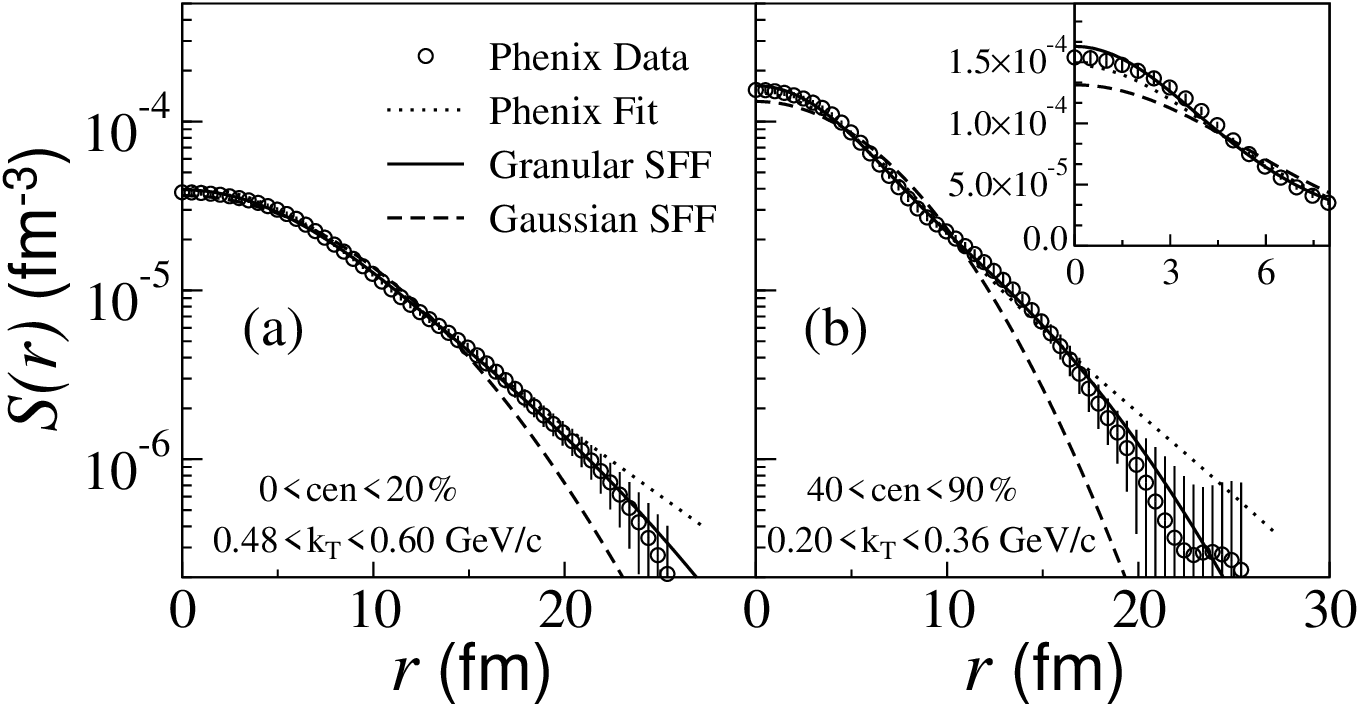}
\caption{\label{fig:fig6} The two-pion source functions ($\circ$
symbols) for $\sqrt{s_{NN}}= 200$ GeV Au+Au collisions \cite{PHE07}
and the SFF curves.}
\end{figure}

\begin{table*}
\caption{\label{tab:table1}The results of granular and Gaussian
source function fits (SFF).}
\begin{ruledtabular}
\begin{tabular}{lccccccc}
&&AGS(a)&AGS(b)&AGS(c)&AGS(d)&RHIC(a)&RHIC(b)\\
\hline &&&&&&&\\
Granular&$R_{\rm gr}$(fm)&$6.31\pm0.23$&$5.92\pm0.18$&$5.28\pm0.19$&$5.39\pm0.23$&$5.23\pm0.21$&$4.53\pm0.10$\\
SFF&$a$(fm)&$2.67\pm0.16$&$2.92\pm0.18$&$2.51\pm0.21$&$2.35\pm0.17$&$3.90\pm0.13$&$2.56\pm0.05$\\
&$\lambda$&$0.93\pm0.05$&$0.77\pm0.02$&$0.57\pm0.02$&$0.65\pm0.04$&$0.24\pm0.01$&$0.39\pm0.01$\\
&$N$&50(fixed)&50(fixed)&50(fixed)&50(fixed)&$3.54\pm0.61$&$4.61\pm0.35$\\
&$\chi^2\!/\,$NDF&0.04&0.05&0.01&0.06&0.35&0.71\\
&$\xi$&$0.28\pm0.08$&$0.17\pm0.05$&$0.19\pm0.07$&$0.25\pm0.09$&$1.56\pm0.97$&$2.12\pm0.55$\\
Gaussian&$R_{\rm ga}$(fm)&$6.26\pm0.11$&$6.22\pm0.07$&$5.53\pm0.09$&$5.41\pm0.12$&$5.05\pm0.13$&$3.79\pm0.03$\\
SFF&$\lambda$&$0.84\pm0.03$&$0.73\pm0.02$&$0.53\pm0.02$&$0.59\pm0.03$&$0.21\pm0.01$&$0.32\pm0.01$\\
&$\chi^2\!/\,$NDF&0.62&0.46&0.32&0.41&3.72&10.87\\
\end{tabular}
\end{ruledtabular}
\end{table*}

From Fig. 6 it can be seen that for the higher $k_{\rm T}$ most
central collisions (a) the source function has not obvious
two-tiered structure and the spheroidal (dot-line) \cite{PHE07},
granular, and Gaussian SFF curves in small $r$ region are almost
overlapped.  It indicates that there is not visible granularity for
the source.  However, for the lower $k_{\rm T}$ peripheral
collisions (b) one can find an obvious two-tiered structure of the
source function.  In small $r$ region the source function has an
enhancement relative to the Gaussian SFF curve [see the insert in
Fig. 6(b)].  Based on the granular source explanation, the
enhancement in small $r$ region indicates that there are small
droplets with higher density in the particle-emitting source.  In
Table I, the value of $\xi$ for the case (b) is larger than that for
the case (a), which is consistent with the observations.  Further
investigation for the reasons of the enhancement of source function
in small $r$ region will be of great interest.

\section{Summary and Conclusion}

We investigated the source imaging for a granular pion-emitting
source model.  The two-pion source functions of the granular sources
exhibit a two-tiered structure, which can be characterized by the
quantity $\xi=(R_{\rm gr}/a)^3/(N-2)$.  In small relative coordinate
$r$ region, the granular two-pion source functions have an
enhancement because of the higher density in the droplets.  We find
that the imaging technique is suitable for probing the granularity
of the pion-emitting sources.  Using a parametrized formula of
granular source function, we examine the two-tiered structure of the
source functions for the imaging data of Au+Au collisions at 2, 4,
6, 8 $A$GeV \cite{E89501} and $\sqrt{s_{NN}}=200$ GeV \cite{PHE07}.
Our analysis results indicate that there is not visible granularity
for the sources produced in the collisions at the AGS energies and
at RHIC energy with the selections $0 < {\rm centrality} < 20\%$ and
$0.48 < k_{\rm T} < 0.60$ GeV/c.  However, the data for the RHIC
collisions with the selections $40 < {\rm centrality} < 90\%$ and
$0.20<k_{\rm T}<0.36$ GeV/c are better described by the model with
granular emission than from that of one Gaussian.  The model with
granular source has more parameters than the simple gaussian, hence
can describe more complicated shapes.

Although our granular parametrized formula of source function is
obtained from a static granular source model and does not include
the effect of droplet overlap, the fitting results with the formula
for evolving sources have still referential meaning.  In this paper
we only examine one-dimension imaging of granular sources.  Because
the longitudinal dynamics at the RHIC energy is very different from
that at AGS energies, the examinations of the source imaging in
different directions \cite{Dan04,Bro05,Dan07,Bro07} and at
intermediate energies ({\it i. e.} SPS energies) will be of great
interest. Further investigations on the source granularity and its
variation with the centrality and particle transverse momentum in
collisions will be also interesting issues.

\begin{acknowledgments}
The authors would like to thank Dr. D. A. Brown for helpful
discussions.  This research was supported by the National Natural
Science Foundation of China under Contracts No. 10575024 and No.
10775024.
\end{acknowledgments}

\end{document}